%% file: iswcs22.tex
\documentclass[conference]{IEEEtran}
\IEEEoverridecommandlockouts

\usepackage{cite}
\usepackage{amsmath,amssymb,amsfonts}
\usepackage{algorithmic}
\usepackage{algorithm}
\usepackage{graphicx}
\usepackage{textcomp}
\usepackage{xcolor}
\usepackage{siunitx}
\usepackage{tikz}
\usepackage{pgfplots}
\usepgfplotslibrary{statistics}
\usetikzlibrary{arrows,shapes}
\usetikzlibrary{decorations.pathmorphing,calc,shapes,shapes.geometric,patterns}
\pgfplotsset{compat=1.8}
\usepackage{caption}

\usepackage{balance}

\newcommand{\plotwidthURAoverSNR}             {0.95\columnwidth}
\newcommand{\plotheightURAoverSNR}            {0.65\columnwidth}

\usepackage{subcaption}

\definecolor{myblack}{RGB}{70,70,70}
\definecolor{myblue}{RGB}{65,105,225}
\definecolor{mygreen}{RGB}{0,139,139}
\definecolor{myorange}{RGB}{255,150,0}
\definecolor{myred}{RGB}{255,69,0}
\definecolor{mylila}{RGB}{153,50,204}

\newcommand{\legendCnn}         {\footnotesize CNN}

\newcommand{\legendGmm}         {\footnotesize GMM}

\newcommand{\legendLs}          {\footnotesize LS}
\newcommand{\legendOmp}         {\footnotesize gen. OMP}

\newcommand{\legendGmmquad}     {\footnotesize GMM (QuaDRiGa)}

\newcommand{\legendGmmtoep}     {\footnotesize Toep. GMM}
\newcommand{\legendGmmcirc}     {\footnotesize circ. GMM}
\newcommand{\legendml}          {\footnotesize ML}
\newcommand{\legendLmmse}       {\footnotesize sample cov.}

\newcommand{\legendtenk}        {\footnotesize $M = 10$}
\newcommand{\legendtwentyfivek} {\footnotesize $M = 25$}
\newcommand{\legendfiftyk}      {\footnotesize $M = 50$}
\newcommand{\legendhundredk}    {\footnotesize $M = 100$}
\newcommand{\legendthreehundk}  {\footnotesize $M = 300$}
\newcommand{\legendfifehundk}   {\footnotesize $M = 500$}

\newcommand{\lineWidth}{1.0pt}
\newcommand{\marksize}{2.0pt}

\tikzset{amp/.style={mark options={solid},color=TUMBeamerYellow, line width=\lineWidth, mark size=\marksize, dashdotted}}
\tikzset{cae/.style={mark options={solid},color=TUMBeamerGreen, line width=\lineWidth, mark=triangle, mark size=\marksize, dashed}}
\tikzset{channelnet/.style={mark options={solid},color=black, line width=\lineWidth, mark=x, mark size=\marksize, dotted}}
\tikzset{cnn/.style={mark options={solid},color=black, line width=\lineWidth, mark=x, mark size=\marksize, dotted}}
\tikzset{genielmmse/.style={mark options={solid},color=TUMBeamerRed, line width=\lineWidth}}
\tikzset{globalcov/.style={mark options={solid},color=TUMBeamerOrange, line width=\lineWidth, mark=o, mark size=\marksize, dashed}}
\tikzset{gmm/.style={mark options={solid},color=TUMBeamerBlue, line width=\lineWidth, mark=square, mark size=\marksize, dashed}}
\tikzset{gmmdiag/.style={mark options={solid},color=mylila, line width=\lineWidth, mark=diamond, mark size=\marksize, dashdotted}}
\tikzset{ls/.style={mark options={solid},color=black, line width=\lineWidth, mark=, mark size=\marksize}}
\tikzset{omp/.style={mark options={solid},color=TUMBeamerGreen, line width=\lineWidth, mark=triangle, mark size=\marksize, dashed}}
\tikzset{gmmcirc/.style={mark options={solid},color=mylila, line width=\lineWidth, mark=diamond, mark size=\marksize, dashdotted}}
\tikzset{gmmquadriga/.style={mark options={solid},color=myred, line width=\lineWidth, mark=pentagon, mark size=\marksize, dashed}}
\tikzset{lmmse/.style={mark options={solid},color=TUMBeamerOrange, mark = o,  line width=\lineWidth, dashed}}

\tikzset{gmm3gpp/.style={mark options={solid},color=TUMBeamerOrange, line width=\lineWidth, mark=o, mark size=\marksize, dashed}}
\tikzset{gmmtoep/.style={mark options={solid},color=TUMMediumGray, line width=\lineWidth, mark=triangle, mark size=\marksize, dashed}}
\tikzset{ml/.style={mark options={solid},color=TUMBeamerYellow, line width=\lineWidth, mark size=\marksize, dashdotted}}

\tikzset{10k/.style={mark options={solid},color=TUMBeamerYellow, line width=\lineWidth, mark size=\marksize, mark=diamond, dashed}}
\tikzset{25k/.style={mark options={solid},color=TUMBeamerOrange, line width=\lineWidth, mark size=\marksize, mark = triangle, dashed}}
\tikzset{50k/.style={mark options={solid},color=mylila, line width=\lineWidth, mark size=\marksize, mark = square, dashed}}
\tikzset{100k/.style={mark options={solid},color=TUMBeamerBlue, line width=\lineWidth, mark size=\marksize, mark = x, dashed}}
\tikzset{300k/.style={mark options={solid},color=TUMBeamerGreen, line width=\lineWidth, mark size=\marksize, mark = o, dashed}}
\tikzset{500k/.style={mark options={solid},color=TUMBeamerRed, line width=\lineWidth, mark size=\marksize, mark =,}}

\def\BibTeX{{\rm B\kern-.05em{\sc i\kern-.025em b}\kern-.08em
    T\kern-.1667em\lower.7ex\hbox{E}\kern-.125emX}}

\input{miko}

\Crefname{figure}{Fig.}{Figs.}

\newacronym{AWGN}{AWGN}{additive white Gaussian noise}
\newacronym{BLMMSE}{BLMMSE}{Bussgang LMMSE}
\newacronym{BS}{BS}{base station}
\newacronym{CDF}{CDF}{cumulative distribution function}
\newacronym{CNN}{CNN}{convolutional neural network}
\newacronym{CSI}{CSI}{channel state information}
\newacronym{CSIT}{CSIT}{channel state information at the transmitter}
\newacronym{DFT}{DFT}{discrete Fourier transform}
\newacronym{DL}{DL}{downlink}
\newacronym{DNN}{DNN}{deep neural network}
\newacronym{DoA}{DoA}{direction of arrival}
\newacronym{EM}{EM}{expectation-maximization}
\newacronym{FDD}{FDD}{frequency division duplex}
\newacronym{GMM}{GMM}{Gaussian mixture model}
\newacronym{LMMSE}{LMMSE}{linear minimum mean square error}
\newacronym{LOS}{LOS}{line of sight}
\newacronym{LS}{LS}{least squares}
\newacronym{MSE}{MSE}{mean squared error}
\newacronym{MIMO}{MIMO}{multiple-input multiple-output}
\newacronym{MPC}{MPC}{multi-path component}
\newacronym{MT}{MT}{mobile terminal}
\newacronym{NLOS}{NLOS}{non-line of sight}
\newacronym{NN}{NN}{neural network}
\newacronym{O2I}{O2I}{outdoor-to-indoor}
\newacronym{OMP}{OMP}{orthogonal matching pursuit}
\newacronym{PDF}{PDF}{probability density function}
\newacronym{PGD}{PGD}{projected gradient descent}
\newacronym{PSD}{PSD}{power spectral density}
\newacronym{SNR}{SNR}{signal-to-noise ratio}
\newacronym{TDD}{TDD}{time division duplex}
\newacronym{UL}{UL}{uplink}
\newacronym{ULA}{ULA}{uniform linear array}
\newacronym{URA}{URA}{uniform rectangular array}
\newacronym{UMa}{UMa}{urban macrocell}
\newacronym{UMi}{UMi}{urban microcell}
\newacronym{nSE}{nSE}{normalized spectral efficiency}
\newacronym{cCDF}{cCDF}{complementary cumulative distribution function}
\newacronym{CME}{CME}{conditional mean estimator}
\newacronym{OFDM}{OFDM}{orthogonal frequency-division multiplexing}
\newacronym{LTE}{LTE}{Long Term Evolution}
\newacronym{GPS}{GPS}{Global Positioning System}
\newacronym{ce}{CE}{channel estimation}
\newacronym{ml}{ML}{machine learning}
\newacronym{phy}{PHY}{physical layer}

\pgfplotsset{compat=1.15}

\newcommand{\Nv}{N_{\mathrm{v}}}
\newcommand{\Nh}{N_{\mathrm{h}}}

\begin{document}



\title{Evaluation of a Gaussian Mixture Model-based Channel Estimator using Measurement Data}


\author{\centerline{Nurettin Turan, Benedikt Fesl, Moritz Grundei, Michael Koller, and Wolfgang Utschick}\\
\IEEEauthorblockA{Professur f\"ur Methoden der Signalverarbeitung, Technische Universit\"at M\"unchen, 80333 Munich, Germany\\
Email: \small{{\{nurettin.turan,benedikt.fesl,moritz.grundei,michael.koller,utschick\}@tum.de}}}
\thanks{The authors acknowledge the financial support by the Federal Ministry of Education and Research of Germany in the program of “Souverän. Digital. Vernetzt.”. Joint project 6G-life, project identification number: 16KISK002}
\thanks{\copyright This work has been submitted to the IEEE for possible publication. Copyright may be transferred without notice, after which this version may no longer be accessible.
}
}

\maketitle

\begin{abstract}

In this work, we use real-world data in order to evaluate and validate a \ac{ml}-based algorithm for physical layer functionalities.
Specifically, we apply a recently introduced \ac{GMM}-based algorithm in order to estimate uplink channels stemming from a measurement campaign.
For this estimator, there is an initial (offline) training phase, where a \ac{GMM} is fitted onto given channel (training) data.
Thereafter, the fitted \ac{GMM} is used for (online) channel estimation.
Our experiments suggest that the \ac{GMM} estimator learns the intrinsic characteristics of a given base station's whole radio propagation environment.
Essentially, this ambient information is captured due to universal approximation properties of the initially fitted \ac{GMM}.
For a large enough number of \ac{GMM} components, the \ac{GMM} estimator was shown to approximate the (unknown) \ac{MSE}-optimal channel estimator arbitrarily well.
In our experiments, the \ac{GMM} estimator shows significant performance gains compared to approaches that are not able to capture the ambient information.
To validate the claim that ambient information is learnt, we generate synthetic channel data using a state-of-the-art channel simulator and train the \ac{GMM} estimator once on these and once on the real data, and we apply the estimator once to the synthetic and once to the real data.
We then observe how providing suitable ambient information in the training phase beneficially impacts the later channel estimation performance.

\end{abstract}


\section{Introduction}

Modern communications systems increasingly utilize \ac{ml} algorithms to meet the compound requirements of high-dimensional \ac{ce} in massive \ac{MIMO} \ac{OFDM} applications \cite{8052521}.
The channel characteristics of the whole propagation environment of a \ac{BS} cell can be described by means of a \ac{PDF} $f_{\mbh}$.
This \ac{PDF} \( f_{\mbh} \) describes the stochastic nature of all channels in the whole coverage area of a \ac{BS} and therefore captures ambient information.
Every channel of any \ac{MT} within the \ac{BS} cell is a realization of a random variable with \ac{PDF} \( f_{\mbh} \).
The main problem is that this \ac{PDF} is typically not available analytically.
For this reason, many classical channel estimation approaches cannot be applied or this ambient information is ignored and replaced with Gaussian assumptions which may only hold locally around a given user.
In this setting, \ac{ml} approaches play an increasingly important role.
These aim to (implicitly) learn the underlying \ac{PDF} from data samples such that the ambient information is taken into account in \ac{ml} channel estimation algorithms \cite{9789120,KoFeTuUt22, KoFeTuUt21J}.

According to this development, many new channel models have been designed to capture the complex channel characteristics of a whole \ac{BS} environment.
Modern channel simulators based on ray tracing or stochastic-geometric models allow for the generation of large synthetic datasets that can be used for training and testing, e.g., \cite{QuaDRiGa1, Alkhateeb2019, Remcom}.
Even though such increasingly complex simulators generate ever more realistic \ac{BS} scenarios, it is crucial to also evaluate the performance of \ac{phy} algorithms on real-world data, i.e., on data collected in a measurement campaign.
This is especially important and interesting for \ac{ml} algorithms which mainly depend on the underlying data (\ac{PDF}), e.g., \cite{9789120,KoFeTuUt21J,TuKoFeBaXuUt}.

An evaluation of an \ac{ml}-based \ac{ce} algorithm on measurement data was done in \cite{HeDeWeKoUt19}, where a neural network-based estimator is analyzed.
However, the estimator in~\cite{HeDeWeKoUt19} was originally derived via assumptions on the channel model and on the antenna configuration which might not hold in practice.
In this work, we evaluate the recently proposed \ac{GMM}-based channel estimator from \cite{KoFeTuUt21J} on data originating from the same measurement campaign.
The estimator first approximates the \ac{PDF} \( f_{\mbh} \) of the whole radio propagation environment with a \ac{GMM}.
This is done offline and only once.
Thereafter, the estimator utilizes this ambient information for \ac{ce} in the online phase.
The estimator is proven to asymptotically converge to the optimal \ac{CME} (which would be calculated using the unknown \ac{PDF} \( f_{\mbh} \)) but so far was only evaluated on synthetic data.


Our experiments indicate that the \ac{GMM} estimator captures the ambient information well
because it outperforms state-of-the-art \ac{ce} algorithms evaluated (and trained) on the same measurement data.
In particular, we achieve lower \acp{MSE} and higher spectral efficiencies.
In addition, we generate synthetic channel data using a state-of-the-art channel simulator and train the \ac{GMM} estimator once on these and once on the measurement data, and we apply the estimator once to the synthetic and once to the measurement data for evaluation.
We observe that providing suitable ambient information in the training phase beneficially impacts the \ac{ce} performance.

The remainder of this work is organized as follows.
In \Cref{sec:ce_gmm}, the \ac{GMM} channel estimator is explained and in \Cref{sec:campaign}, the measurement campaign is described and a channel simulator is introduced for comparison. 
\Cref{sec:sim_results} provides simulation results and \Cref{sec:conlusion} concludes this work.

\section{Gaussian Mixture Model Channel Estimator}
\label{sec:ce_gmm}

We consider \ac{ce} in the uplink from a single-antenna \ac{MT} located within the cell to a \ac{BS}. 
The \ac{BS} is equipped with $N$ antennas.
After correlating with the commonly known pilot sequence, we obtain the noisy observation
\begin{equation}
    \mby = \mbh + \mbn \in \C^{N},
\end{equation}
where $\mbh \in \C^{N}$ is the uplink-channel of a certain \ac{MT} located within the coverage area of the \ac{BS} and $\mbn \sim \mathcal{N}_\C(\mathbf{0}, \mbC_{\mbn} = \sigma^2 \mbI)$ denotes the \ac{AWGN}.
The goal is then to estimate the channel $\mbh$ given $\mby$, i.e., to denoise the observation $\mby$.
The stochastic nature of all channels in the whole coverage area of the \ac{BS} is assumed to be described by means of a continuous \ac{PDF} \( f_{\mbh} \).
Every channel \( \mbh \) of any \ac{MT} located within the \ac{BS} cell is a realization of a random variable with \ac{PDF} \( f_{\mbh} \).
For such a system model, the \ac{MSE}-optimal channel estimator is given by the \ac{CME}
\begin{equation}\label{eq:conditional_mean}
    \hhat = \expec[\mbh \mid \mby] = \int \mbh f_{\mbh\mid\mby}(\mbh\mid\mby) d \mbh,
\end{equation}
which can generally not be computed analytically.
Further, \( f_{\mbh} \) is typically not available in an analytic form.

However, in \cite{KoFeTuUt21J} a method to approximate~\eqref{eq:conditional_mean} with the help of \acp{GMM} was proposed.
To this end, assuming to have access to a set \( \mathcal{H}_M = \{ \mbh_m \}_{m=1}^M \) of training channel samples, which represent the radio propagation environment (ambient information), and motivated by universal approximation properties of \acp{GMM}~\cite{NgNgChMc20}, we fit a \ac{GMM} \( f_{\mbh}^{(K)} \) with \( K \) components to \( \mathcal{H}_M \) in order to approximate the unknown channel \ac{PDF} \( f_{\mbh} \).

A \ac{GMM} is a \ac{PDF} of the form~\cite{bookBi06}
\begin{equation}\label{eq:gmm_of_h}
    f^{(K)}_{\mbh}(\mbh) = \sum_{k=1}^K p(k) \calN_{\C}(\mbh; \mbmu_k, \mbC_k),
\end{equation}
where every summand is one of its \( K \) components.
It is characterized by the means $\mbmu_k \in \C^N$, the covariances $\mbC_k \in \C^{N\times N}$, and the mixing coefficients $p(k)$. Maximum likelihood estimates of these parameters can be computed using an \ac{EM} algorithm and the training data set \( \mc{H}_M \), see~\cite{bookBi06}.


The idea in~\cite{KoFeTuUt21J} is to compute the \ac{MSE}-optimal estimator \( \hhat_{\text{GMM}}^{(K)} \) for channels distributed according to \( f_{\mbh}^{(K)} \) and to use it to estimate the channels distributed according to \( f_{\mbh} \).
This estimator \( \hhat_{\text{GMM}}^{(K)} \) converges pointwise to the optimal estimator \( \hhat \) from~\eqref{eq:conditional_mean} as \( K \to \infty \), cf.~\cite{KoFeTuUt21J}.

Once the (offline) \ac{GMM} fitting process is done,
the (online) channel estimates
can be computed in closed form:
\begin{equation}\label{eq:gmm_estimator_closed_form}
    \hhat_{\text{GMM}}^{(K)}(\mby) = \sum_{k=1}^K p(k \mid \mby) \hhat_{\text{LMMSE},k}(\mby),
\end{equation}
with the responsibilities
\begin{equation}\label{eq:responsibilities}
    p(k \mid \mby) = \frac{p(k) \calN_{\C}(\mby; \meanhk, \covhk + \mbC_{\mbn})}{\sum_{i=1}^K p(i) \calN_{\C}(\mby; \meanhi, \covhi + \mbC_{\mbn}) },
\end{equation}
and
\begin{equation}\label{eq:lmmse_formula}
    \hhat_{\text{LMMSE},k}(\mby) =
    \covhk ( \covhk + \mbC_{\mbn})^{-1} (\mby -  \meanhk) + \meanhk.
\end{equation}
The weights \( p(k \mid \mby) \) are the probabilities that component \( k \) generated the current observation \( \mby \), cf.~\cite{KoFeTuUt21J}.

\subsection{Complexity Analysis and Low Cost Adaptations}
The inverse in \eqref{eq:lmmse_formula} can be precomputed offline for various \acp{SNR} because the \ac{GMM} covariance matrices \( \covhk \) do not change once the \ac{GMM} fitting process is done.
Accordingly, evaluating \eqref{eq:lmmse_formula} online is dominated by matrix-vector multiplications and has a complexity of \( \mc{O}(N^2) \).
It remains to calculate the responsibilities~\eqref{eq:responsibilities} by evaluating Gaussian densities.
A Gaussian density with mean \( \mbmu \in \C^N \) and covariance matrix \( \mbC \in \C^{N\times N} \) can be written as
\begin{equation}\label{eq:gaussian_density}
    \calN_{\C}(\mbh; \mbmu, \mbC) = \frac{\exp(-(\mbh - \mbmu)^\herm \mbC^{-1} (\mbh - \mbmu))}{\pi^N \det(\mbC)}.
\end{equation}
Again, since the \ac{GMM} covariance matrices and mean vectors do not change between observations, the inverses and the determinants of the densities can be pre-computed offline.
Therefore, the online evaluation is also in this case dominated by matrix-vector multiplications and has a complexity of \( \calO(N^2) \).
Overall, evaluating \eqref{eq:gmm_estimator_closed_form} has a complexity of \( \calO(K N^2) \) \cite{KoFeTuUt21J}.
Since \( \hhat_{\text{GMM}}^{(K)} \) converges pointwise to the \ac{MSE}-optimal estimator $\hat{\mbh}$ form \eqref{eq:conditional_mean} as \( K \to \infty \),
a trade-off between the performance of the estimator and the complexity can be achieved by adjusting the number \( K \) of \ac{GMM} components. 

The complexity of the estimator from \eqref{eq:gmm_estimator_closed_form} can be reduced by introducing structural constraints to the \ac{GMM} covariance matrices \( \covhk \).
For example, in case of a \ac{ULA} employed at the \ac{BS}, it is common to assume Toeplitz covariance matrices, see, e.g., \cite{NeWiUt18}.
Further, for large numbers of antenna elements, a Toeplitz matrix is well approximated by a circulant matrix~\cite{Gr06}.
Motivated by these common assumptions we enforce structural constraints onto the \ac{GMM} covariances.
Since we consider exclusively an environment, where at the \ac{BS} a \ac{URA} with $\Nv$ vertical and $\Nh$ horizontal ($N=\Nv \times \Nh$) elements is employed, the structural assumptions result in block-Toeplitz matrices with Toeplitz blocks, or block-circulant matrices with circulant blocks, respectively \cite{KoHeUt19}.
In general the structured covariances can be expressed as
\begin{equation}
    \covhk = \mbQ^\herm \diag(\mbc_k) \mbQ,
\end{equation}
where on the one hand, when assuming a Toeplitz structure, $\mbQ = \mbQ_{\Nv} \otimes \mbQ_{\Nh}$, where $\mbQ_J$ contains the first $J$ colums of a $2J\times 2J$ \ac{DFT} matrix, and $\mbc_k \in \R_{+}^{4\Nh\Nv}$ \cite{KoHeUt19, St86}.
On the other hand, when assuming circular structure, we have $\mbQ = \mbF_{\Nv} \otimes \mbF_{\Nh}$, where $\mbF_J$ is the $J\times J$ DFT-matrix, and $\mbc_k \in \R_{+}^{\Nh\Nv}$.
In both cases, the structural constraints allow to store only the $\mbc_k$'s of a \ac{GMM} which drastically reduces the memory overhead and the number of parameters to be learned, similar as in \cite{KoFeTuUt21J, FeJoHuKoTuUt22}.

Further, in case of circular covariances, the complexity of evaluating \eqref{eq:gmm_estimator_closed_form} reduces to \( \calO(K N \log(N)) \), where 2D-DFT transforms are exploited when evaluating~\eqref{eq:responsibilities} and~\eqref{eq:lmmse_formula}, cf. \cite{KoFeTuUt21J}.

\section{Measurement Campaign and Synthetic Data}
\label{sec:campaign}
\begin{figure}
    \centering
    \begin{tikzpicture}
        \draw (0,0) node[below right] {\includegraphics[width=0.975\columnwidth]{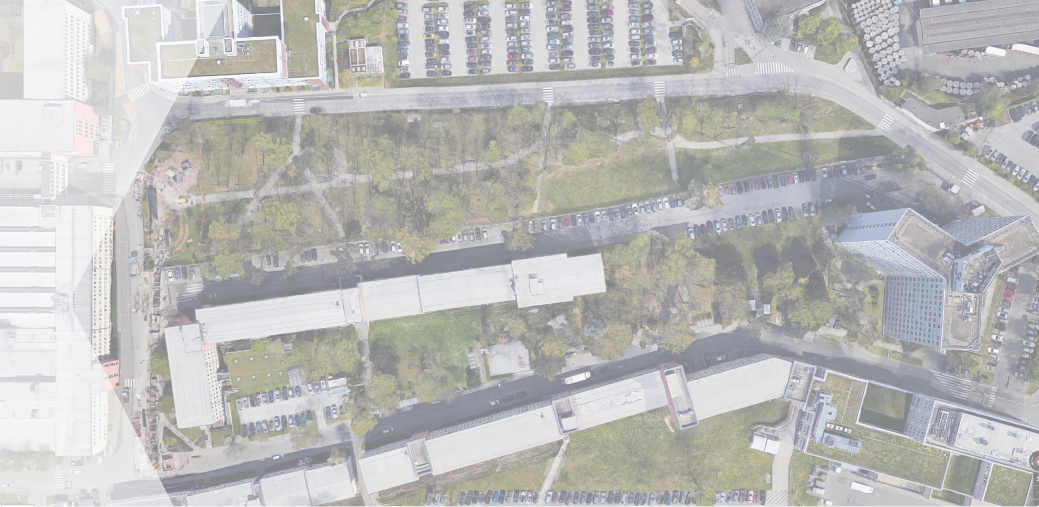}};
        \draw (0,0) node[below right] {\includegraphics[width=0.975\columnwidth]{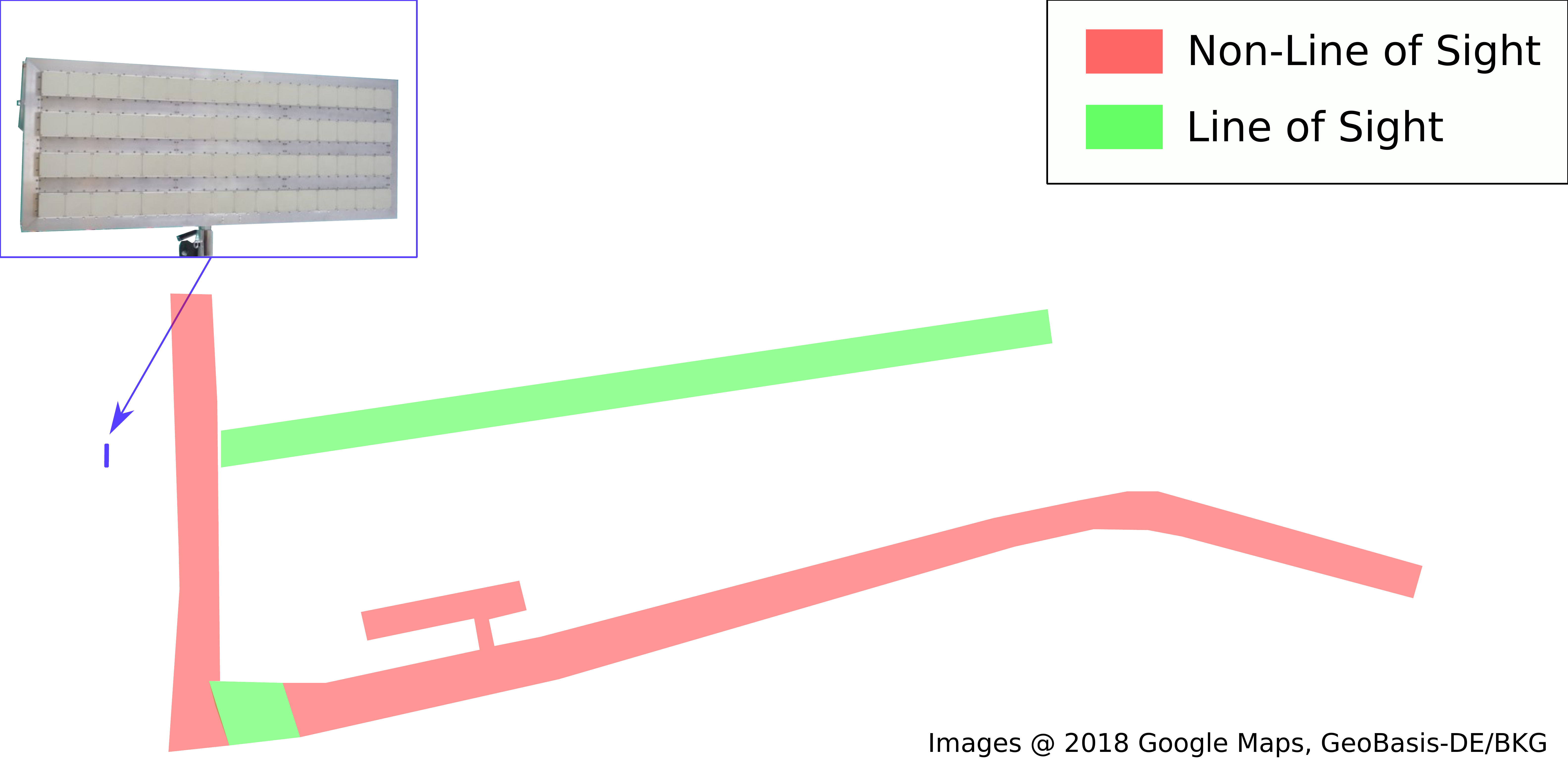}};
    \end{tikzpicture}
    \caption{Measurement setup on the Nokia campus in Stuttgart, Germany.}
    \label{fig:meas_campaign}
\end{figure}

The measurement campaign was conducted at the Nokia campus in Stuttgart, Germany, in October/November 2017.
As can be seen in \Cref{fig:meas_campaign}, the receive antenna with a down-tilt of $\SI{10}{\degree}$ was mounted on a rooftop about $\SI{20}{m}$ above the ground and comprises a \ac{URA} with $\Nv=4$ vertical and $\Nh=16$ horizontal single polarized patch antennas.
The horizontal spacing is $\lambda$ and the vertical spacing equals $\lambda/2$, where the geometry of the \ac{BS} antenna array was adapted to the \ac{UMi} propagation scenario, which exhibited a larger horizontal than vertical angular spread.
The carrier frequency is $\SI{2.18}{\giga\hertz}$.
The \ac{BS} transmitted time-frequency orthogonal pilots using $\SI{10}{\mega\hertz}$ \ac{OFDM} waveforms.
In particular, $600$ sub-carriers with $\SI{15}{\kilo\hertz}$ spacing were used, which resembles typical \ac{LTE} numerology.
The pilots were sent continuously with a periodicity of $\SI{0.5}{ms}$ and were arranged in $50$ separate subbands, with $12$ consecutive subcarriers each, for channel sounding purposes.
For the duration of one pilot burst the propagation channel was assumed to remain constant.

The single monopole receive antenna, which mimics the \ac{MT}, was mounted on top of a moving vehicle at a height of $\SI{1.5}{m}$.
The maximum speed was $\SI{25}{kmph}$.
Synchronization between the transmitter and receiver was achieved using GPS.
The data was collected by a TSMW receiver and stored on a Rohde \& Schwarz IQR hard disk recorder.

In a post-processing step, by the correlation of the received signal with the pilot sequence a channel realization vector with $N=\Nv \times \Nh$ coefficients per subband was extracted.
The measurement was conducted at a high \ac{SNR}, which ranged from $\SI{20}{dB}$ to $\SI{30}{dB}$.
Thus, the measured channels are regarded as ground truth.
Further, we assume fully calibrated antennas and thus channel reciprocity is assumed. 
In this work, we will therefore consider a system, where we artificially corrupt the measured channels with \ac{AWGN} at specific \acp{SNR} and thereby obtain noisy observations $\mby = \mbh + \mbn$.
The task is then to denoise the observations and obtain an estimated channel $\hat{\mbh}$.
We want to highlight that we investigate a single-snapshot scenario, i.e., the
coherence interval of the covariance matrix and of the channel is identical (the channel covariance matrix changes at the
same time scale as the channel).

\subsection{Synthetic Data Generation using QuaDRiGa}

Version $2.6.1$ of the QuaDRiGa channel simulator \cite{QuaDRiGa1, QuaDRiGa2} was used to generate \ac{CSI} in a \ac{UMi} scenario.
The environment for which synthetic data is generated was adapted as closely as possible to the circumstances of the measurement environment.
For this reason, the carrier frequency was set to $\SI{2.18}{\giga\hertz}$.
The base station is placed at a height of 20 meters.
The minimum and maximum distances between \acp{MT} and the \ac{BS} are $\SI{35}{\meter}$ and $\SI{315}{\meter}$, respectively.
The \acp{MT} are located outdoors at a height of $\SI{1.5}{\meter}$.
A QuaDRiGa channel is given by \( \mbh = \sum_{\ell=1}^{L} \mbg_{\ell} e^{-2\pi j f_c \tau_{\ell}} \)
with $\ell$ the path number, $L$ the number of \acp{MPC}, $f_c$ the carrier frequency, and $\tau_{\ell}$ the \( \ell \)th path delay.
The number \( L \) depends on whether there is \ac{LOS} or \ac{NLOS} propagation, cf. \cite{QuaDRiGa2}.
The coefficients vector \( \mbg_{\ell} \) consists of one complex entry for each antenna pair and comprises the attenuation of a path, the antenna radiation pattern weighting, and the polarization.
As described in the QuaDRiGa manual~\cite{QuaDRiGa2}, the generated channels are post-processed to remove the path gain.

\section{Experiments and Results}
\label{sec:sim_results}

Normalizing the data so that $\expec[\|\mbh\|^2] = N$ allows us to define an \ac{SNR} in our simulations as \( \frac{1}{\sigma^2} \).
Given $T$ test samples \( \{ \mbh_t \}_{t=1}^T \) and obtaining corresponding channel estimates \( \{ \hhat_t \}_{t=1}^T \), we use the normalized \ac{MSE} \( \frac{1}{NT} \sum_{t=1}^T \| \mbh_t - \hhat_t \|^2\) as performance measure.

The following baseline estimators are considered.
In the system at hand, the least squares estimate is simply given by the noisy observations \( \hhat_{\text{LS}} = \mby \).
Another baseline is the sample covariance matrix based approach, where we construct a sample covariance matrix $\mbC_s = \frac{1}{M} \sum_{m=1}^M \mbh_m \mbh_m^\herm$ given a set of training samples and calculate \ac{LMMSE} channel estimates $\hhat_{\text{s-cov}} = \mbC_s (\mbC_s + \mbC_{\mbn})^{-1} \mby$.


Compressive sensing approaches commonly assume that the channel exhibits a certain structure: $\mbh \approx \mbD \mbt$, where $\mbD \in \C^{N\times L}$ is a dictionary. We used an oversampled \ac{DFT} matrix as dictionary, with $L=4N$ (cf., e.g., \cite{AlLeHe15}).
A compressive sensing algorithm like \ac{OMP}~\cite{Gharavi} can then be used to obtain a sparse vector \( \mbt \), and the estimated channel is calculated as $\hhat_{\text{OMP}} = \mbD \mbt$. Since the sparsity order of the channel is not known but the algorithm's performance crucially depends on it, we use a genie-aided approach to obtain a bound on the performance of the algorithm. In particular, we use the true channel to choose the optimal sparsity order.

We further compare to a \ac{CNN}-based channel estimator, which was introduced in~\cite{NeWiUt18}.
There, the authors exploit assumptions which stem from a spatial channel model (3GPP, cf.~\cite{3GPP_scm}) in order to derive the \ac{CNN} architecture. The \ac{CNN} is then trained on measurement data to compensate the mismatch of the assumptions and the real world data. We use the rectified linear unit as activation function and the input transform is based on the \( 2N \times 2N \) \ac{DFT} matrix, cf. \cite[Equation (43)]{NeWiUt18}.

\input{simresultstex/simomeasURAOctoverSNR}

In \Cref{fig:meas_URA_over_snr_NMSE}, we use $T=10{,}000$ channel samples stemming from the measurement campaign for evaluating the performances of the different channel estimators.
In particular, we compare the GMM estimator with full covariances, denoted by ``GMM'', and block Toeplitz (``Toep. GMM'') or block circulant (``circ. GMM'') covariances, to the state-of-the-art estimators described above. 
In total we use $M=300{,}000$ channel samples stemming from the measurement campaign as training data in the fitting process of the \ac{GMM} approaches, each with $K=64$ components.
Also the learning process of the \ac{CNN} estimator (``CNN'') and the construction of a sample covariance matrix for the sample covariance \ac{LMMSE} estimation approach (``sample cov.'') use these \( M \) samples.

The \ac{GMM} estimator with full covariance matrices performs best over the whole SNR range from $\SI{-15}{dB}$ to $\SI{20}{dB}$, followed by the ``Toep. GMM'' and the ``circ. GMM'' approaches.
As expected, the GMM estimator’s performance suffers from introducing structural covariance constraints but it still outperforms the other channel estimation approaches.
With ``GMM (QuaDRiGa)'' we depict the \ac{GMM} approach where synthetic training data ($M=300{,}000$ samples) is used to fit the \ac{GMM}. 
Despite using synthetic data of an environment, which was adapted as closely as possible to the circumstances of the measurement campaign's environment, we can observe a severe performance degradation in the estimation performance of the \ac{GMM} estimator.

In \Cref{fig:quad_URA_over_snr_NMSE}, we  replace the test data and use $T=10{,}000$ synthetic channel samples for comparing the ``GMM'' estimator (fitted with measurement data), the ``CNN'' approach (trained on measurement data), the ``sample cov.'' approach (sample covariance obtained using measurement data) with the ``GMM (QuaDRiGa)'' (fitted with synthetic data) approach.
We can observe that the ``GMM (QuaDRiGa)'' approach now performs best since the learned ambient information now matches the synthetic test data on which the estimator is evaluated.
We conclude that using synthetic channel data is not suitable to replicate the 
ambient information of the campus where the measurement campaign was conducted, and vice versa.
Thus, this validates the claim that ambient information is learnt by the \ac{GMM} when provided suitable training data.
We further want to highlight that the \ac{CNN} estimator which constitutes a data based approach as well, should also be able to capture the underlying ambient information of the considered propagation environment. 
Up to some extent this seems to be the case since the \ac{CNN} approach exhibits the best performance right after the \ac{GMM} approaches. 
Nevertheless, even the \ac{GMM} approach with block circulant covariances, which has the same order of complexity as the \ac{CNN} approach, yields a better overall estimation performance.

\input{simresultstex/simonquadURAOctoberSNR}

In \Cref{fig:meas_URA_over_snr_DR} we consider the same simulation parameters as in \Cref{fig:meas_URA_over_snr_NMSE}, and analyze a performance upper bound for the achievable spectral efficiency \begin{equation}
    \bar{r} = E\left[\log_2\left(1 + \dfrac{|\hat{\bm{h}}^H\bm{h}|^2}{\sigma^2||\hat{\bm{h}}||^2}\right)\right],
\end{equation}
when applying a matched filter $\frac{\hat{\bm{h}}^H}{||\hat{\bm{h}}||}$ in the uplink \cite{NeWiUt18}, which may also be interpreted as a measure of the accuracy of the estimated channel subspace \cite{NeWiUt18}.
Note, that there is no one-by-one relation between the spectral efficiency and the \ac{MSE} in general, cf. \cite{HeDeWeKoUt19}.
In essence, we can observe that the \ac{GMM} yields the best performance while there is only a minor degradation when using structured covariances.

\input{simresultstex/simDRonmeasURAOctoberSNR}

\input{simresultstex/simCompareSamples}
\input{simresultstex/simCompareSamplesToepCirc}

In \Cref{fig:comp_samples}, we depict the behavior of the \ac{GMM} estimator with full covariances for a varying number of components $K$ and a varying number of training data used to fit the \ac{GMM}. 
The \ac{SNR} is $\SI{10}{dB}$.
The number of parameters of a \ac{GMM} increases with an increasing $K$, which requires more training data to achieve a good fit.
As the figure suggests, as long as there are enough training data,
increasing $K$ leads to a better performance.

In contrast, in \Cref{fig:compare_samples_toep_circ}, where we consider \acp{GMM} with fewer parameters by assuming either block Toeplitz (top) or block circulant (bottom) covariances, we can observe that the estimator already achieves a good performance with a low to moderate number of available training samples. 
In particular, with structured covariances, increasing $K$ leads to a better performance already with more than $100{,}000$ available training samples.
Overall, in \Cref{fig:comp_samples} and \Cref{fig:compare_samples_toep_circ}, even a small to moderate number of components ($K=16$ or $K=32$) performs well. Accordingly, a suitable number of components $K$ should be determined based on the amount of available training data and the desired overall estimation complexity.

\input{simresultstex/simomeas_comps_valmeas}
\input{simresultstex/simomeas_comps_valquad}



In \Cref{fig:avg_resps_meas} and \Cref{fig:avg_resps_quad}, we aim to investigate the differences in the distributions of the synthetic and the measurement data from a different perspective:
We plot the average responsibilities, cf. \eqref{eq:responsibilities}, of the \ac{GMM} when fitted on either synthetic (``GMM (QuaDRiGa)'') or on measurement (``GMM'') data.
To this end, we evaluate  \( p(k \mid \mby) \) for each observation $\mby$ and average these responsibilities with respect to the samples in the evaluation set with $T=10{,}000$ samples and an SNR of $\SI{10}{dB}$.
Afterwards, we sort the components from highest to lowest average responsibility.
In \Cref{fig:avg_resps_meas}, the evaluation set contains only measurement data. 
It can be observed that the average responsibilities of the \ac{GMM} fitted on synthetic or on measurement data are similar up to some extent.
That is, the channel simulator is able to capture general information about the underlying \ac{UMi} scenario---but not the details of the measurement environment in its full extent.
In contrast, in \Cref{fig:avg_resps_quad}, the same \acp{GMM} are evaluated with synthetic data.
The average responsibilities can be clearly distinguished since a large mismatch can be observed.
A possible explanation for this observation is that the \ac{GMM} fitted onto the measurement data is specifically designed for the environment of the measurement campaign with unique immanent characteristics.
In contrast to the channel simulator with a stochastic nature (hence aiming to model general \ac{UMi} scenarios), this \ac{GMM} does not generalize to different \ac{UMi} scenarios.
This behavior is desirable since the distinctive design allows for performance gains as discussed earlier.
\Cref{fig:avg_resps_meas} and \Cref{fig:avg_resps_quad} show above all that the attempt to represent channel data with the false \ac{GMM} only works to a limited extent, which can be seen, among other performance metrics, from the fact that fewer components of the false \ac{GMM} are identified as representative than would be the case with the correct \ac{GMM}.
This underlines the importance of an evaluation with measurement data.



\section{Conclusion and Outlook}
\label{sec:conlusion}
In this work, we used real-world data stemming from a measurement campaign in order to evaluate and validate a recently introduced \ac{GMM}-based algorithm for uplink channel estimation.
Our experiments suggest that the \ac{GMM} estimator learns the intrinsic characteristics of a given base station's whole radio propagation environment.
To validate the claim that ambient information is learnt, we conducted experiments, where we used test data either stemming from the measurement campaign or synthetic data.
We observed that providing suitable ambient information, which is implicitly contained within the data, in the training phase (offline), beneficially impacts the channel estimation performance in the online phase.
We further showed that structurally constrained covariances of the GMM, which are motivated by model-based insights, also work well when using real-world data. 
In particular, one can drastically reduce the computational complexity and memory overhead with only small performance losses.
An immediate additional advantage is that less training data is needed due to the lower number of \ac{GMM} parameters, which need to be fitted, when assuming structural constraints.
Future work might consider a more accurate and involved emulation of the propagation environment using a digital twin. For example, a digital representative of the propagation environment can be generated using a ray tracing tool, where the measurement campus with all of the buildings and streets, which are characteristic for certain propagation properties, is recreated virtually. 
Given the digital twin of the propagation environment, the performance of the data based channel estimators might be evaluated under these more accurate digital representatives.

\balance
\bibliographystyle{IEEEtran}
\bibliography{IEEEabrv,biblio}
\vfill

\end{document}

%% file: miko.tex
\usepackage{mathtools}	
\usepackage{amssymb}
\usepackage{amsthm}

\usepackage[utf8]{inputenc}

\usepackage{ifthen}

\usepackage{microtype}

\usepackage{caption}

\usepackage{bm}

\usepackage[draft,nomargin,marginclue,inline]{fixme}
\makeatletter
\renewcommand*\FXLayoutMarginClue[3]{%
  \marginpar[%
  \raggedleft\@fxuseface{margin}\textcolor{red}{\ignorespaces $ \Rightarrow $}]{%
    \raggedright\@fxuseface{margin}\textcolor{red}{\ignorespaces $ \Leftarrow $}}}
\makeatother	

\usepackage{tumcolor}

\usepackage{pgfplotstable}	

\pgfplotsset{
	discard if/.style 2 args={
        x filter/.append code={
            \edef\tempa{\thisrow{#1}}
            \edef\tempb{#2}
            \ifx\tempa\tempb
                
            \fi
        }
    },
    discard if not/.style 2 args={
        x filter/.append code={
            \edef\tempa{\thisrow{#1}}
            \edef\tempb{#2}
            \ifx\tempa\tempb
            \else
                
            \fi
        }
    }
}

\usepackage{cite}

\usepackage[acronym,shortcuts]{glossaries}
\newacronym{cnn}{CNN}{convolutional neural network}
\newacronym{ula}{ULA}{uniform linear array}

\usepackage{cleveref}

\usepackage{enumitem}

\tikzset{algorithm1/.style={mark options={solid},color=TUMBeamerBlue, line width=\lineWidth, mark=square, dashed}}

\pgfplotscreateplotcyclelist{miko}{%
{color=TUMBeamerRed, dash pattern=on 5pt off 1pt on 1pt off 1pt, line width=1.5pt},%
{color=TUMBeamerOrange, dash pattern=on 5pt off 2pt, line width=1.5pt,
mark=o},%
{color=black, dash pattern=on 5pt off 2pt on 3pt off 2pt, line width=1.5pt},%
{color=TUMDarkerBlue, line width=1.5pt},%
{color=black, dotted, line width=1.5pt},%
{color=TUMBeamerGreen, dotted, line width=1.5pt},%
{color=TUMBeamerYellow, dotted, line width=1.5pt},%
{color=TUMBeamerDarkRed, dotted, line width=1.5pt},%
{color=TUMBeamerLightGreen, dotted, line width=1.5pt},%
{color=TUMIvory, dotted, line width=1.5pt},%
{color=TUMLightBlue, dotted, line width=1.5pt},%
}

\DeclareMathOperator{\diag}{diag}

\DeclareMathOperator{\expec}{E}

\newcommand*{\mc}[1]{\mathcal{#1}}	

\newcommand{\calN}{\mathcal{N}}
\newcommand{\calO}{\mathcal{O}}

\newcommand*{\C}{\mathbb{C}}

\newcommand*{\R}{\mathbb{R}}

\newcommand{\herm}{{\operatorname{H}}}

\definecolor{myblue}{RGB}{30, 100, 200}

\newlength{\leftstackrelawd}
\newlength{\leftstackrelbwd}
\def\leftstackrel#1#2{\settowidth{\leftstackrelawd}%
	{${{}^{#1}}$}\settowidth{\leftstackrelbwd}{$#2$}%
	\addtolength{\leftstackrelawd}{-\leftstackrelbwd}%
	\leavevmode\ifthenelse{\lengthtest{\leftstackrelawd>0pt}}%
	{\kern-.5\leftstackrelawd}{}\mathrel{\mathop{#2}\limits^{#1}}}

\newcommand{\mbC}{\bm{C}}
\newcommand{\mbD}{\bm{D}}

\newcommand{\mbF}{\bm{F}}

\newcommand{\mbI}{\bm{I}}

\newcommand{\mbQ}{\bm{Q}}

\newcommand{\mbc}{\bm{c}}

\newcommand{\mbg}{\bm{g}}
\newcommand{\mbh}{\bm{h}}

\newcommand{\mbn}{\bm{n}}

\newcommand{\mbt}{\bm{t}}

\newcommand{\mby}{\bm{y}}

\newcommand{\mbmu}{{\bm{\mu}}}

\newcommand{\hhat}{\hat{\mbh}}

\newcommand{\covhi}{\mbC_i}
\newcommand{\covhk}{\mbC_k}
\newcommand{\meanhi}{\mbmu_i}
\newcommand{\meanhk}{\mbmu_k}

%% file: simresultstex/simomeasURAOctoverSNR.tex
    \begin{figure}[t]
	\centering
	\begin{tikzpicture}
		\begin{axis}
			[width=\plotwidthURAoverSNR,
			height=\plotheightURAoverSNR,
			xtick=data,
			xmin=-15, 
			xmax=20,
			xlabel={SNR [$\SI{}{dB}$]},
			ymode = log, 
			ymin= 5e-3,
			ymax=1,
			ylabel= {Normalized MSE}, 
			ylabel shift = 0.0cm,
			grid = both,
			legend columns = 3,
			legend entries={
                \legendLs,
				\legendCnn,
				\legendGmmquad,
				\legendGmm,
				\legendGmmtoep,
				\legendGmmcirc,
				\legendOmp,
				\legendLmmse,
			},
			legend style={at={(0.5,1.05)}, anchor=south},
			legend cell align = {left},
			]
			\addplot[ls]
			table[x=SNR in dB, y=LS, col sep=comma]
			{csvdat/comparison_estimators_NMSE_Oct_URA_val_meas_64_comp.csv};
	
			\addplot[cnn]
			table[x=SNR in dB, y=CNN, col sep=comma]
			{csvdat/comparison_estimators_NMSE_Oct_URA_val_meas_64_comp.csv};

			\addplot[gmmquadriga]
			table[x=SNR in dB, y=GMMquad, col sep=comma]
			{csvdat/comparison_estimators_NMSE_Oct_URA_val_meas_64_comp.csv};

			\addplot[gmm]
			table[x=SNR in dB, y=GMMmeas, col sep=comma]
			{csvdat/comparison_estimators_NMSE_Oct_URA_val_meas_64_comp.csv};

			\addplot[gmmtoep]
			table[x=SNR in dB, y=GMMtoep, col sep=comma]
			{csvdat/comparison_estimators_NMSE_Oct_URA_val_meas_64_comp.csv};

			\addplot[gmmcirc]
			table[x=SNR in dB, y=GMMcirc, col sep=comma]
			{csvdat/comparison_estimators_NMSE_Oct_URA_val_meas_64_comp.csv};
			
			\addplot[omp]
			table[x=SNR in dB, y=GenieOmp, col sep=comma]
			{csvdat/comparison_estimators_NMSE_Oct_URA_val_meas_64_comp.csv};

			
			\addplot[lmmse]
			table[x=SNR in dB, y=LMMSE, col sep=comma]
			{csvdat/comparison_estimators_NMSE_Oct_URA_val_meas_64_comp.csv};
			
		\end{axis}
	\end{tikzpicture}
	\caption{Normalized MSE for various estimators over the SNR (evaluated on measurement data, with $T=10{,}000$ samples). Each GMM approach is constructed using $K=64$ components.}
    \label{fig:meas_URA_over_snr_NMSE}
\end{figure}
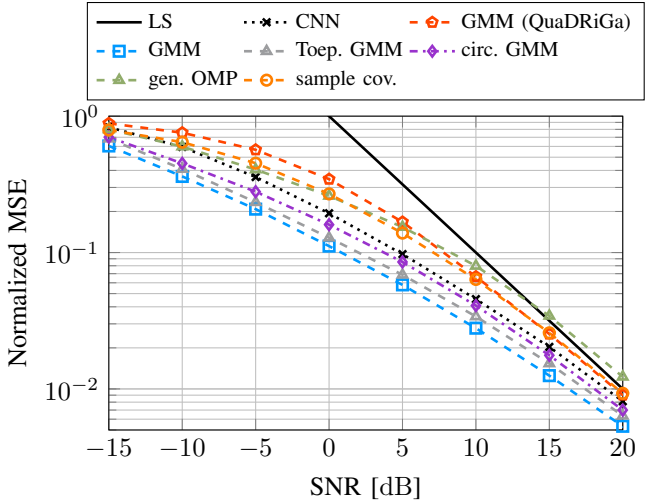

%% file: simresultstex/simonquadURAOctoberSNR.tex
    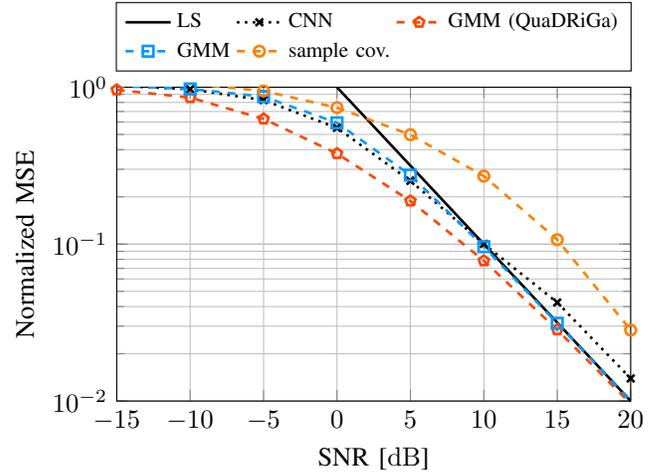
\begin{figure}[t]
	\centering
	\begin{tikzpicture}
		\begin{axis}
			[width=\plotwidthURAoverSNR,
			height=\plotheightURAoverSNR,
			xtick=data,
			xmin=-15, 
			xmax=20,
			xlabel={SNR [$\SI{}{dB}$]},
			ymode = log, 
			ymin= 1e-2,
			ymax=1.0,
			ylabel= {Normalized MSE}, 
			ylabel shift = 0.0cm,
			grid = both,
			legend columns = 3,
			legend entries={
			    \legendLs,
				\legendCnn,
				\legendGmmquad,
				\legendGmm,
				\legendLmmse,
			},
			legend style={at={(0.5,1.05)}, anchor=south},
			legend cell align = {left},
			]
			\addplot[ls]
			table[x=snr, y=LS, col sep=comma]
			{csvdat/comparison_estimators_NMSE_Oct_URA_val_quad_64_comp.csv};
			
			\addplot[cnn]
			table[x=snr, y=CNN, col sep=comma]
			{csvdat/comparison_estimators_NMSE_Oct_URA_val_quad_64_comp.csv};

			\addplot[gmmquadriga]
			table[x=snr, y=GMMquad, col sep=comma]
			{csvdat/comparison_estimators_NMSE_Oct_URA_val_quad_64_comp.csv};

			\addplot[gmm]
			table[x=snr, y=GMMreal, col sep=comma]
			{csvdat/comparison_estimators_NMSE_Oct_URA_val_quad_64_comp.csv};




			
			\addplot[lmmse]
			table[x=snr, y=LMMSE, col sep=comma]
			{csvdat/comparison_estimators_NMSE_Oct_URA_val_quad_64_comp.csv};

		\end{axis}
	\end{tikzpicture}
	\caption{Normalized MSE for various estimators over the SNR (evaluated on synthetic data, with $T=10{,}000$ samples). Each GMM approach is constructed using $K=64$ components.}
    \label{fig:quad_URA_over_snr_NMSE}
\end{figure}

%% file: simresultstex/simDRonmeasURAOctoberSNR.tex
    \begin{figure}[t]
	\centering
	\begin{tikzpicture}
		\begin{axis}
			[width=\plotwidthURAoverSNR,
			height=\plotheightURAoverSNR,
			xtick=data,
			xmin=-15, 
			xmax=10,
			xlabel={SNR [$\SI{}{dB}$]},
			ymode = log, 
			ymin= 1e-1,
			ymax = 10,
			ylabel= {spectral efficiency [bits/c.u.]}, 
			ylabel shift = 0.0cm,
			grid = both,
			legend columns = 3,
			legend entries={
                \legendLs,
				\legendCnn,
				\legendGmmquad,
				\legendGmm,
				\legendGmmtoep,
				\legendGmmcirc,
				\legendOmp,
				\legendLmmse,
			},
			legend style={at={(0.5,1.05)}, anchor=south},
			legend cell align = {left},
			]
	        \addplot[ls]
			table[x=snr, y=LS, col sep=comma]
			{csvdat/comparison_estimators_DR_Oct_URA_val_meas_64_comp.csv};
			
			\addplot[cnn]
			table[x=snr, y=CNN, col sep=comma]
			{csvdat/comparison_estimators_DR_Oct_URA_val_meas_64_comp.csv};

			\addplot[gmmquadriga]
			table[x=snr, y=GMMquad, col sep=comma]
			{csvdat/comparison_estimators_DR_Oct_URA_val_meas_64_comp.csv};

			\addplot[gmm]
			table[x=snr, y=GMMreal, col sep=comma]
			{csvdat/comparison_estimators_DR_Oct_URA_val_meas_64_comp.csv};

			\addplot[gmmtoep]
			table[x=snr, y=GMMtoep, col sep=comma]
			{csvdat/comparison_estimators_DR_Oct_URA_val_meas_64_comp.csv};

            \addplot[gmmcirc]
			table[x=snr, y=GMMcirc, col sep=comma]
			{csvdat/comparison_estimators_DR_Oct_URA_val_meas_64_comp.csv};

			\addplot[omp]
			table[x=snr, y=GenieOmp, col sep=comma]
			{csvdat/comparison_estimators_DR_Oct_URA_val_meas_64_comp.csv};


			\addplot[lmmse]
			table[x=snr, y=LMMSE, col sep=comma]
			{csvdat/comparison_estimators_DR_Oct_URA_val_meas_64_comp.csv};
	
		\end{axis}
	\end{tikzpicture}
	\caption{Achievable spectral efficiencies using various estimators over the SNR (evaluated on measurement data, with $T=10{,}000$ samples). Each GMM approach is constructed using $K=64$ components.}
    \label{fig:meas_URA_over_snr_DR}
\end{figure}
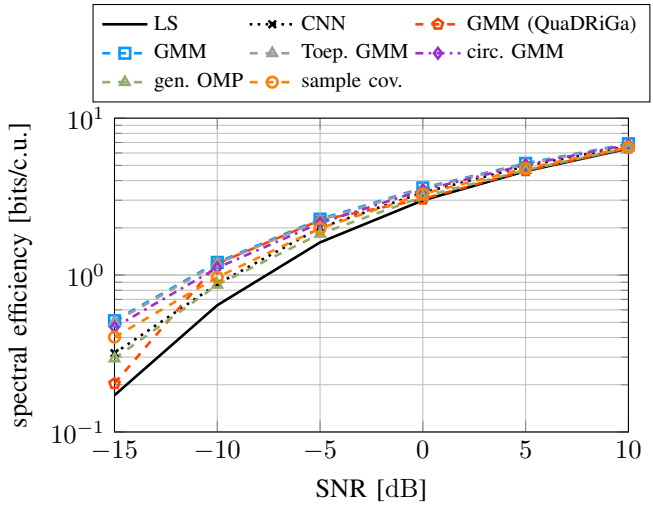

%% file: simresultstex/simCompareSamples.tex
\begin{figure}[t]
	\centering
	\begin{tikzpicture}
		\begin{axis}
			[width=\plotwidthURAoverSNR,
			height=\plotheightURAoverSNR,
			xtick=data,
			xmin=1, 
			xmax=128,
			xlabel={\( K \), number of GMM components},
			ymode = log, 
			ymin= 2e-2,
			ymax=3e-1,
			ytick={1e-1},
			extra y ticks={3e-2,4e-2,5e-2,6e-2,7e-2,8e-2,9e-2},
			extra y tick labels={$3\cdot 10^{-2}$, ,$5\cdot 10^{-2}$},
			ylabel= {Normalized MSE},
			extra y tick style={
                log identify minor tick positions=true},
			ylabel shift = 0.0cm,
			grid = both,
			legend columns = 3,
			legend entries={
				\legendtenk,
                \legendtwentyfivek,
                \legendfiftyk,
                \legendhundredk,
                \legendthreehundk,
                \legendfifehundk
			},
			legend style={at={(0.5,1.05)}, anchor=south},
			legend cell align = {left},
			]

			\addplot[10k]
			table[x=components, y=10k, col sep=comma]
			{csvdat/compare_samples.csv};

			\addplot[25k]
			table[x=components, y=25k, col sep=comma]
			{csvdat/compare_samples.csv};

			\addplot[50k]
			table[x=components, y=50k, col sep=comma]
			{csvdat/compare_samples.csv};

			\addplot[100k]
			table[x=components, y=100k, col sep=comma]
			{csvdat/compare_samples.csv};

			\addplot[300k]
			table[x=components, y=300k, col sep=comma]
			{csvdat/compare_samples.csv};

			\addplot[500k]
			table[x=components, y=500k, col sep=comma]
			{csvdat/compare_samples.csv};
	
		\end{axis}
	\end{tikzpicture}
	\caption{Normalized MSE of the GMM estimator with full covariance matrices over the number of components $K$ (evaluated on measurement data, with $T=10{,}000$ samples). The GMM estimator is fitted using $M \cdot 10^3$ samples. The SNR is $\SI{10}{dB}$.}
    \label{fig:comp_samples}
\end{figure}
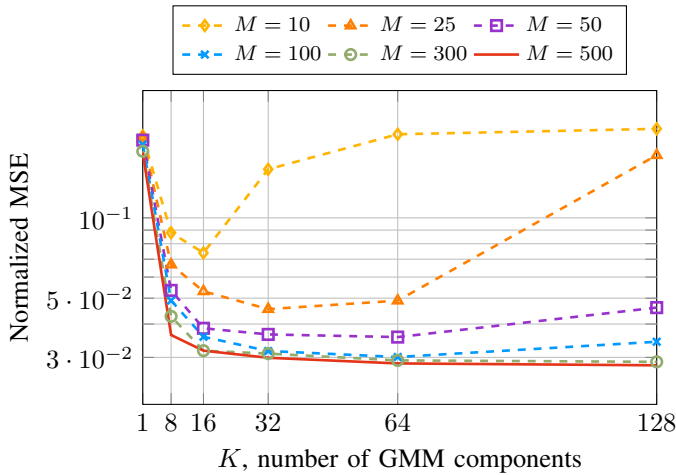

%% file: simresultstex/simCompareSamplesToepCirc.tex
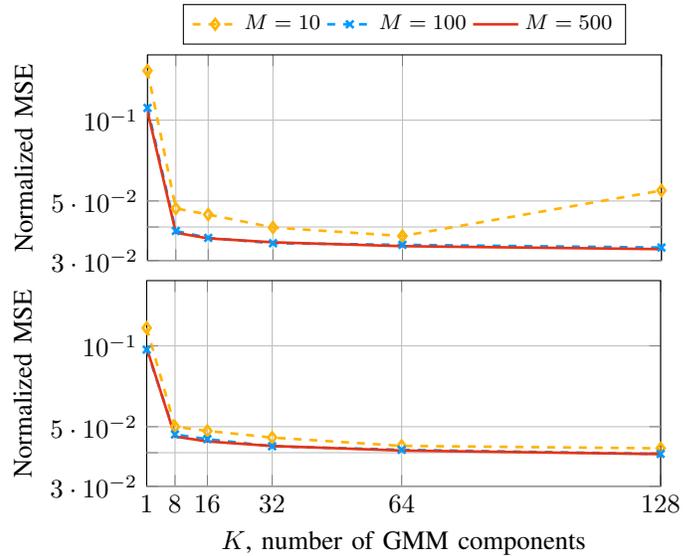
\begin{figure}[t]
    \centering
    \input{simresultstex/simCompareSamplesToep}
    \input{simresultstex/simCompareSamplesCirc}
    \begin{tikzpicture} 
        \begin{axis}[%
            hide axis,
            xmin=10,
            xmax=50,
            ymin=0,
            ymax=0.4,
            legend style={draw=white!15!black,legend cell align=left},
            legend columns = 5,
        	legend entries={
        	},
        	legend style={at={(0.0,1.0)}, anchor=south west},
        	legend cell align = {left}
            ]
        \end{axis}
    \end{tikzpicture}
    \caption{Normalized MSE of the GMM estimator with block Toeplitz (top) or block circulant (bottom) covariances over the number of components $K$ (evaluated on measurement data, with $T=10{,}000$ samples). The respective GMM estimator is fitted using $M \cdot 10^3$ samples. The SNR is $\SI{10}{dB}$.
    }
    \label{fig:compare_samples_toep_circ}
\end{figure}

%% file: simresultstex/simCompareSamplesToep.tex
\begin{subfigure}[b]{1\columnwidth}
	\centering
	\begin{tikzpicture}
		\begin{axis}
			[width=\plotwidthURAoverSNR,
			height=0.75*\plotheightURAoverSNR,
			xtick=data,
			xmin=1, 
			xmax=128,
			xticklabels = {},
			ymode = log, 
			ymin= 3e-2,
			ymax=1.75e-1,
			ytick={1e-1},
			extra y ticks={3e-2,4e-2,5e-2},
			extra y tick labels={$3\cdot 10^{-2}$, ,$5\cdot 10^{-2}$},
			ylabel= {Normalized MSE},
			extra y tick style={
                log identify minor tick positions=true},
			ylabel shift = 0.0cm,
			grid = both,
			legend columns = 3,
			legend entries={
				\legendtenk,
                \legendhundredk,
                \legendfifehundk
			},
			legend style={at={(0.5,1.05)}, anchor=south},
			legend cell align = {left},
			]

			\addplot[10k]
			table[x=components, y=10k, col sep=comma]
			{csvdat/compare_samples_toep.csv};



			\addplot[100k]
			table[x=components, y=100k, col sep=comma]
			{csvdat/compare_samples_toep.csv};


			\addplot[500k]
			table[x=components, y=500k, col sep=comma]
			{csvdat/compare_samples_toep.csv};
	
		\end{axis}
	\end{tikzpicture}
\end{subfigure}

%% file: simresultstex/simCompareSamplesCirc.tex
\begin{subfigure}[b]{1\columnwidth}
	\centering
	\begin{tikzpicture}
		\begin{axis}
			[width=\plotwidthURAoverSNR,
			height=0.75*\plotheightURAoverSNR,
			xtick=data,
			xmin=1, 
			xmax=128,
			xtick=data,
			xlabel={\( K \), number of GMM components},
			ymode = log, 
			ymin= 3e-2,
			ymax=1.75e-1,
			ytick={1e-1},
			extra y ticks={3e-2,4e-2,5e-2},
			extra y tick labels={$3\cdot 10^{-2}$, ,$5\cdot 10^{-2}$},
			ylabel= {Normalized MSE},
			extra y tick style={
                log identify minor tick positions=true},
			ylabel shift = 0.0cm,
			grid = both,
			]

			\addplot[10k]
			table[x=components, y=10k, col sep=comma]
			{csvdat/compare_samples_circ.csv};



			\addplot[100k]
			table[x=components, y=100k, col sep=comma]
			{csvdat/compare_samples_circ.csv};


			\addplot[500k]
			table[x=components, y=500k, col sep=comma]
			{csvdat/compare_samples_circ.csv};
	
		\end{axis}
	\end{tikzpicture}
\end{subfigure}

%% file: simresultstex/simomeas_comps_valmeas.tex
    \begin{figure}[t]
	\centering
		\begin{tikzpicture}
		\begin{axis}
			[ybar=0pt,
			bar width=0.45pt,
			line width=0.05pt,
			log origin=infty,
			width=0.9\columnwidth,
			height=0.52\columnwidth,
			xtick={1,16,32,48,64,80,96,112,128},
			xmin=0, 
			xmax=128,
			xlabel={GMM component},
			ymode = log, 
			ymin= 1e-5,
			ymax=1,
			ylabel= {avg. Responsibility}, 
			ylabel shift = 0.0cm,
			ytick={1, 1e-1, 1e-2, 1e-3, 1e-4, 1e-5},
			grid = major,
			tick label style={font=\small},
	    	label style={font=\small},
			legend columns = 2,
			legend entries={
				GMM (QuaDRiGa),
				GMM,
			},
			legend style={font=\scriptsize,at={(1.0,1.0)}, anchor=north east},
			]
			
			\addplot[color=TUMBeamerRed,line width=0.2pt,pattern=north east lines, pattern color=TUMBeamerRed]
			table[x=components, y=quad, col sep=comma]
			{csvdat/responsibilities_val_meas.csv};
			
			\addplot[mark options={solid},color=TUMBeamerBlue,line width=0.2pt,fill=TUMBeamerBlue,opacity=1.0,pattern=north west lines, pattern color=TUMBeamerBlue, style=solid]
			table[x=components, y=meas, col sep=comma]
			{csvdat/responsibilities_val_meas.csv};
			
		\end{axis}
	\end{tikzpicture}
	\caption{Average responsibilities of the GMM components (evaluated on measurement data, with $T=10{,}000$ samples). The SNR is $\SI{10}{dB}$.}
    \label{fig:avg_resps_meas}
\end{figure}
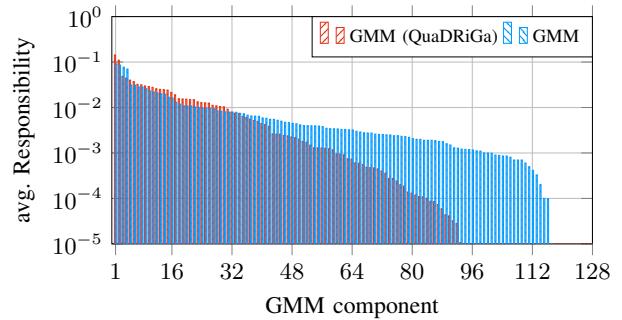

%% file: simresultstex/simomeas_comps_valquad.tex
    \begin{figure}[t]
	\centering
		\begin{tikzpicture}
		\begin{axis}
			[ybar=0pt,
			bar width=0.45pt,
			line width=0.05pt,
			log origin=infty,
			width=0.9\columnwidth,
			height=0.52\columnwidth,
			xtick={1,16,32,48,64,80,96,112,128},
			xmin=0, 
			xmax=128,
			xlabel={GMM component},
			ymode = log, 
			ymin= 1e-5,
			ymax=1,
			ylabel= {avg. Responsibility}, 
			ylabel shift = 0.0cm,
			ytick={1, 1e-1, 1e-2, 1e-3, 1e-4, 1e-5},
			grid = major,
			tick label style={font=\small},
	    	label style={font=\small},
			legend columns = 2,
			legend entries={
				GMM (QuaDRiGa),
				GMM,
			},
			legend style={font=\scriptsize,at={(1.0,1.0)}, anchor=north east},
			]
			
			\addplot[color=TUMBeamerRed,line width=0.2pt,pattern=north east lines, pattern color=TUMBeamerRed]
			table[x=component, y=quad, col sep=comma]
			{csvdat/responsibilities_val_quad.csv};
			
			\addplot[mark options={solid},color=TUMBeamerBlue,line width=0.2pt,fill=TUMBeamerBlue,opacity=1.0,pattern=north west lines, pattern color=TUMBeamerBlue, style=solid]
			table[x=component, y=meas, col sep=comma]
			{csvdat/responsibilities_val_quad.csv};
			
		\end{axis}
	\end{tikzpicture}
	\caption{Average responsibilities of the GMM components (evaluated on synthetic data, with $T=10{,}000$ samples). The SNR is $\SI{10}{dB}$.}
    \label{fig:avg_resps_quad}
\end{figure}
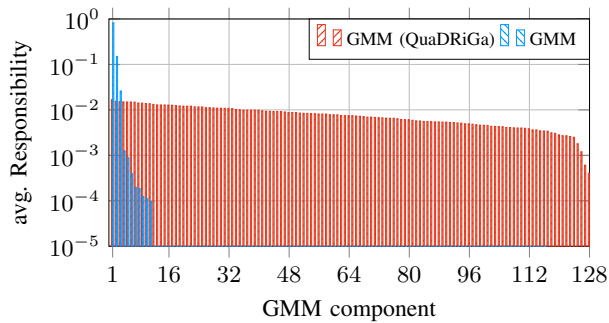

%% file: iswcs22.bbl
\begin{thebibliography}{10}
\providecommand{\url}[1]{#1}
\csname url@samestyle\endcsname
\providecommand{\newblock}{\relax}
\providecommand{\bibinfo}[2]{#2}
\providecommand{\BIBentrySTDinterwordspacing}{\spaceskip=0pt\relax}
\providecommand{\BIBentryALTinterwordstretchfactor}{4}
\providecommand{\BIBentryALTinterwordspacing}{\spaceskip=\fontdimen2\font plus
\BIBentryALTinterwordstretchfactor\fontdimen3\font minus
  \fontdimen4\font\relax}
\providecommand{\BIBforeignlanguage}[2]{{%
\expandafter\ifx\csname l@#1\endcsname\relax
\typeout{** WARNING: IEEEtran.bst: No hyphenation pattern has been}%
\typeout{** loaded for the language `#1'. Using the pattern for}%
\typeout{** the default language instead.}%
\else
\language=\csname l@#1\endcsname
\fi
#2}}
\providecommand{\BIBdecl}{\relax}
\BIBdecl

\bibitem{8052521}
H.~Ye, G.~Y. Li, and B.-H. Juang, ``Power of deep learning for channel
  estimation and signal detection in {OFDM} systems,'' \emph{IEEE Wireless
  Commun. Letters}, vol.~7, no.~1, pp. 114--117, 2018.

\bibitem{9789120}
J.~Guo, C.-K. Wen, M.~Chen, and S.~Jin, ``Environment knowledge-aided massive
  {MIMO} feedback codebook enhancement using artificial intelligence,''
  \emph{{IEEE} Trans. Commun.}, 2022.

\bibitem{KoFeTuUt22}
M.~Koller, B.~Fesl, N.~Turan, and W.~Utschick, ``An asymptotically optimal
  approximation of the conditional mean channel estimator based on {Gaussian}
  mixture models,'' in \emph{IEEE Int. Conf. on Acoust., Speech and Signal
  Process.}, 2022, pp. 5268--5272.

\bibitem{KoFeTuUt21J}
------, ``An asymptotically {MSE}-optimal estimator based on {Gaussian} mixture
  models,'' \emph{{IEEE} Trans. Signal Process.}, pp. 1--14, 2022.

\bibitem{QuaDRiGa1}
S.~{Jaeckel}, L.~{Raschkowski}, K.~{Börner}, and L.~{Thiele}, ``Quadriga: A
  3-d multi-cell channel model with time evolution for enabling virtual field
  trials,'' \emph{{IEEE} Trans. Antennas Propag.}, vol.~62, no.~6, pp.
  3242--3256, 2014.

\bibitem{Alkhateeb2019}
A.~Alkhateeb, ``{DeepMIMO}: A generic deep learning dataset for millimeter wave
  and massive {MIMO} applications,'' in \emph{Proc. of Inf. Theory and Appl.
  Workshop (ITA)}, San Diego, CA, Feb 2019.

\bibitem{Remcom}
Remcom, ``{Wireless InSite},'' \url{http://www.remcom.com/wireless-insite}.

\bibitem{TuKoFeBaXuUt}
N.~Turan, M.~Koller, B.~Fesl, S.~Bazzi, W.~Xu, and W.~Utschick, ``{GMM}-based
  codebook construction and feedback encoding in {FDD} systems,'' 2022, arXiv
  preprint: 2205.12002.

\bibitem{HeDeWeKoUt19}
C.~Hellings, A.~Dehmani, S.~Wesemann, M.~Koller, and W.~Utschick, ``Evaluation
  of neural-network-based channel estimators using measurement data,'' in
  \emph{Proc. Int. ITG Workshop on Smart Antennas (WSA)}, 2019, pp. 1--5.

\bibitem{NgNgChMc20}
T.~T. Nguyen, H.~D. Nguyen, F.~Chamroukhi, and G.~J. McLachlan, ``Approximation
  by finite mixtures of continuous density functions that vanish at infinity,''
  \emph{Cogent Math. Statist.}, vol.~7, no.~1, p. 1750861, 2020.

\bibitem{bookBi06}
C.~M. Bishop, \emph{Pattern Recognition and Machine Learning (Information
  Science and Statistics)}.\hskip 1em plus 0.5em minus 0.4em\relax Berlin,
  Heidelberg: Springer-Verlag, 2006.

\bibitem{NeWiUt18}
D.~Neumann, T.~Wiese, and W.~Utschick, ``Learning the {MMSE} channel
  estimator,'' \emph{{IEEE} Trans. Signal Process.}, vol.~66, no.~11, pp.
  2905--2917, 2018.

\bibitem{Gr06}
R.~M. Gray, ``Toeplitz and circulant matrices: {A} review,'' \emph{Found. and
  Trends\textsuperscript{\textregistered} in Commun. and Inf. Theory}, no.~3,
  pp. 155--239, 2006.

\bibitem{KoHeUt19}
M.~Koller, C.~Hellings, and W.~Utschick, ``Learning-based channel estimation
  for various antenna array configurations,'' in \emph{Proc. IEEE Int. Workshop
  on Signal Process. Advances in Wireless Commun. (SPAWC)}, 2019, pp. 1--5.

\bibitem{St86}
G.~Strang, ``\BIBforeignlanguage{en}{A proposal for {Toeplitz} matrix
  calculations},'' \emph{\BIBforeignlanguage{en}{Stud. in Appl. Math.}},
  vol.~74, no.~2, pp. 171--176, 1986.

\bibitem{FeJoHuKoTuUt22}
B.~Fesl, M.~Joham, S.~Hu, M.~Koller, N.~Turan, and W.~Utschick, ``Channel
  estimation based on {Gaussian} mixture models with structured covariances,''
  2022, arXiv preprint: 2205.03634.

\bibitem{QuaDRiGa2}
S.~{Jaeckel}, L.~{Raschkowski}, K.~{Börner}, L.~{Thiele}, F.~{Burkhardt}, and
  E.~{Eberlein}, ``Quadriga: Quasi deterministic radio channel generator, user
  manual and documentation,'' Fraunhofer Heinrich Hertz Institute, Tech. Rep.,
  v2.2.0, 2019.

\bibitem{AlLeHe15}
A.~Alkhateeb, G.~Leus, and R.~W. Heath, ``Compressed sensing based multi-user
  millimeter wave systems: {How} many measurements are needed?'' in \emph{2015
  IEEE Int. Conf. on Acoust., Speech and Signal Process. (ICASSP)}, 2015, pp.
  2909--2913.

\bibitem{Gharavi}
M.~Gharavi-Alkhansari and T.~Huang, ``A fast orthogonal matching pursuit
  algorithm,'' in \emph{Proc. IEEE Int. Conf. on Acoust., Speech and Signal
  Process. (ICASSP)}, vol.~3, 1998, pp. 1389--1392.

\bibitem{3GPP_scm}
3GPP, ``Spatial channel model for multiple input multiple output ({MIMO})
  simulations,'' 3rd Generation Partnership Project (3GPP), Tech. Rep. 25.996
  (V16.0.0), Jul. 2020.

\end{thebibliography}
